\documentclass[preprint,12pt]{elsarticle}

\usepackage{amssymb}

\journal{Nuclear Physics B}
\usepackage{color}

\usepackage{latexsym}
\usepackage{txfonts}
\input{psfig.sty}

\usepackage{graphicx}

\usepackage{color}

\usepackage{algorithm}
\usepackage{algorithmic}

\newtheorem{Def}{Definition}
\newtheorem{Exm}{Example}

\begin{document}

\begin{frontmatter}



\title{Detecting Spurious Counterexamples Efficiently in Abstract Model Checking\tnoteref{label1}}

\tnotetext[label1]{This research is supported by the NSFC Grant No.
61003078, 91018010, 61133001 and 60910004, 973 Program Grant No.
2010CB328102, SRFDP Grant No. 200807010012 and ISN Lab Grant No.
ISN1102001.}

\author{Cong Tian and Zhenhua Duan\corref{cor1}
}

\cortext[cor1]{Corresponding author. E-mail address:
zhenhua\_duan@126.com}

\address{ICTT and ISN Laboratory, Xidian
University, Xi'an, 710071, P.R. China}

\begin{abstract}
Abstraction is one of the most important strategies for dealing with
the state space explosion problem in model checking. In the abstract
model, the state space is largely reduced, however, a counterexample
found in such a model may not be a real counterexample in the
concrete model. Accordingly, the abstract model needs to be further
refined. How to check whether or not a reported counterexample is
spurious is a key problem in the abstraction-refinement loop. In
this paper, a formal definition for spurious path is given. Based on
it, efficient algorithms for detecting spurious counterexamples are
proposed.

\end{abstract}


\end{frontmatter}



\section{Introduction}
Model checking is an important approach for the verification of
hardware, software, multi-agent systems, communication protocols,
embedded systems and so forth. The term model checking was coined by
Clarke and Emerson \cite{Clarke81}, as well as Sifakis and Queille
\cite{QS82}, independently. The earlier model checking algorithms
explicitly enumerated the reachable states of the system in order to
check the correctness of a given specification. This restricted the
capacity of model checkers to systems with a few million states.
Since the number of states can grow exponentially in the number of
variables, early implementations were only able to handle small
designs and did not scale to examples with industrial complexity. To
combat this, kinds of methods, such as abstraction, partial order
reduction, OBDD, symmetry and bound technique are applied to model
checking to reduce the state space for efficient verification.
Thanks to these efforts, model checking has been one of the most
successful verification approaches which is widely adopted in
industrial community.

Among the techniques for reducing the state space, abstraction is
certainly the most important one. Abstraction technique preserves
all the behaviors of the concrete system but may introduce behaviors
that are not present originally. Thus, if a property (i.e. a
temporal logic formula) is satisfied in the abstract model, it will
still be satisfied in the concrete model. However, if a property is
unsatisfiable in the abstract model, it may still be satisfied in
the concrete model, and none of the behaviors that violate the
property in the abstract model can be reproduced in the concrete
model. In this case, the counterexample is said to be spurious.
Thus, when a spurious counterexample is found, the abstraction
should be refined in order to eliminate the spurious behaviors. This
process is repeated until either a real counterexample is found or
the abstract model satisfies the property.

In the abstraction-refinement loop, how to check whether or not a
reported counterexample is spurious is a key problem. In
\cite{CGJLV00}, algorithm {\scshape SplitPath} is presented for
checking whether or not a counterexample is spurious, and a SAT
solver is employed to implement it \cite{Clarke04,CGHS02}. In
{\scshape SplitPath}, whether or not a counterexample is spurious
can be checked by detecting the first failure state in the
counterexample. If a failure state is found, the counterexample is
spurious, otherwise, the counterexample is a real one. However,
whether or not a state, say $\hat{s_i}$, is a failure state relies
on the prefix of the counterexample
$\hat{s_0},\hat{s_1},...,\hat{s_i}$. This brings in a polynomial
number of unwinding of the loop in an infinite counterexample
\cite{CGJLV00,CGJLV002}.

In this paper, based on a formal definition of failure states,
spurious paths are re-analyzed, and a new approach for checking
spurious counterexamples is proposed. Within this approach, whether
or not a counterexample is spurious still depends on the existence
of failure states in the counterexample. Instead of the prefix, to
checking whether or not a state $\hat{s_i}$ is a failure state is
only up to $\hat{s_i}$'s pre- and post- states in the
counterexample. Based on this, for an infinite counterexample, the
polynomial number of unwinding of the loop can be avoided. Further,
the algorithm can be easily improved by detecting the heaviest
failure state such that a number of model checking iterations can be
saved in the whole abstract-refinement loop. In addition, the
algorithm can be naturally parallelled.

The rest parts of the paper are organized as follows. The next
section briefly presents the preliminaries in
abstraction-refinement. In section 3, why spurious counterexamples
occur is analyzed intuitively and algorithm {\scshape SplitPath} is
briefly presented. In section 4, a formal definition of spurious
counterexamples is given with respect to the formal definition of
failure states. Further, in section 5, efficient algorithms for
checking whether or not a counterexample in the abstract model is
spurious are presented. Finally, conclusions are drawn in section 6.

\section{Abstraction and Refinement}

There are many techniques for obtaining the abstract models
\cite{Rushby99,Krushan94,HJMS02}. We follow the counterexample
guided abstraction and refinement method proposed by Clarke, etc,
where abstraction is performed by selecting a set of variables which
are insensitive to the desired property to be invisible
\cite{Clarke04}. We use $h: S\rightarrow \hat{S}$ to denote an
abstract function, where $S$ is the set of all states in the
original model, and $\hat{S}$ the set of all states in the abstract
model. For clearance,  $s$, $s_1$, $s_2$, ... are usually used to
denote the states in the original model, and $\hat{s}$, $\hat{s_1}$,
$\hat{s_2}$, ... indicate the states in the abstract model. Further,
for a state $\hat{s}$ in the abstract model, $h^-(\hat{s})$ is used
to denote the set of origins of $\hat{s}$ in the original model.

\begin{figure}[htp]
\centerline{\includegraphics[height=3.8cm]{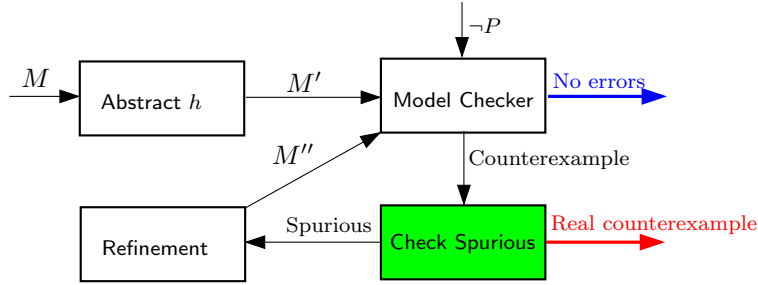}}\caption{Abstraction
refinement loop}\label{fig:Arloop}
\end{figure}

The abstraction-refinement loop is depicted in Fig.\ref{fig:Arloop}.
Initially, the abstract model $M'$ is obtained by the abstract
function $h$. Then a model checker is employed to check whether or
not the abstract model satisfies the desired property. If no errors
are found, the model is correct. Otherwise, a counterexample is
reported and rechecked by a checker which is used to check whether
or not a counterexample is spurious. If the counterexample is not
spurious, it will be a real counterexample that violates the system;
otherwise, the counterexample is spurious, and  a refining tool is
used to refine the abstract model
\cite{CGJLV00,Clarke04,HSHGS10,HSNH10,WLJHS06,GKMFV03}.
Subsequently, the refined abstract model is checked with the model
checker again until either a real counterexample is found or the
model is checked to be correct. In this paper, we concentrate on the
how to check whether or not a counterexample is spurious.

\section{Spurious Paths}
To check a spurious counterexample efficiently, we first show why
spurious paths occur intuitively with an example. Then we briefly
present the basic idea of algorithm {\scshape SplitPath} which is
used in \cite{CGJLV00,CGJLV002} for checking whether or not a
counterexample is spurious.
\subsection{Why Spurious Paths?}
Abstraction technique preserves all the behaviors of the concrete
system but may introduce behaviors that are not present originally.
Therefore, when implementing the model checker with the abstract
model, some reported counterexamples will not be real
counterexamples that violate the desired property. This is
intuitively illustrated by the traffic lights controller example
\cite{CGJLV00}.
\begin{Exm}\rm
For the traffic light controller in Fig. \ref{fig:light} (1), by
making variable \emph{color} invisible, an abstract model can be
obtained as shown in Fig. \ref{fig:light} (2). We want to prove
$\Box\Diamond (state=stop)$ (any time, the state of the light will
be $stop$ sometimes in the future). By implementing model checking
with the abstract model,
\begin{figure}[htp]
\centerline{\includegraphics[height=3.8cm]{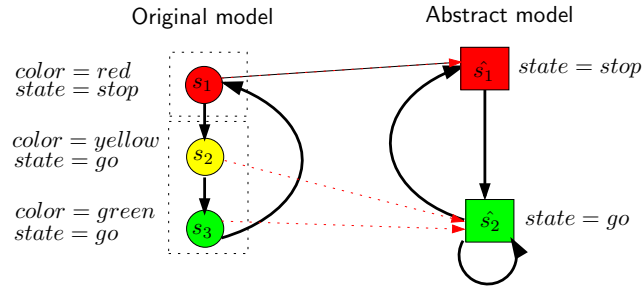}}\caption{Traffic
Light Controller}\label{fig:light}
\end{figure}
 a counterexample, $\hat{s_1},\hat{s_2},\hat{s_2},\hat{s_2},...$ will be reported. However, in the concrete model, such a behavior cannot be found. So, this is not a real counterexample. \hfill{$\Box$}
\end{Exm}

\subsection{Detecting Spurious Counterexample with {\scshape SplitPath}}
\begin{figure}[htp]
\centerline{\includegraphics[height=3.1cm]{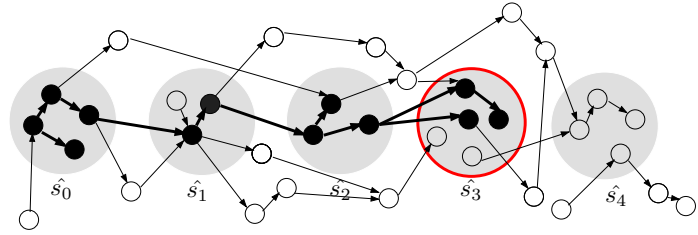}}\caption{Algorithm
{\scshape SplitPath} }\label{fig:split}
\end{figure}

In \cite{CGJLV00}, algorithms {\scshape SplitPath} is presented for
checking whether or not a finite counterexample is spurious. In
{\scshape SplitPath}, as illustrated in Fig.\ref{fig:split},
initially, the set, $M_0$, of starting states falling into
$h^-(\hat{s_0})$,
$$M_0=I\cap h^-(\hat{s_0})$$ is computed.
Then for the image of the states in $I\cap h^-(\hat{s_0})$, i.e.
$R(I\cap h^-(\hat{s_0}))$, the set of states falling into
$h^-(\hat{s_1})$,
 $$M_1=M_0\cap
h^-(\hat{s_1})=R(I\cap h^-(\hat{s_1}))\cap h^-(\hat{s_2})$$ is
computed. Generally, for any $i\geq 1$,

$$
\begin{array}{llll}
M_i&=&R(M_{i-1})\cap h^-(\hat{s_i})\\
&=&R(R(M_{i-2})\cap h^-(\hat{s_{i-1}}))\cap h^-(\hat{s_i})\\
&=&R(R(R(M_{i-1})\cap h^-(\hat{s_i}))\cap h^-(\hat{s_{i-1}}))\cap h^-(\hat{s_i})\\
&=&...\\
&=&R(R(...(I\cap h^-(\hat{s_1}))\cap...\cap h^-(\hat{s_{i-1}}))\cap h^-(\hat{s_i})\\

\end{array}$$
is computed
recursively. For some state $\hat{s_k}$, $k\geq 1$, if
$M_k=\emptyset$, $\hat{s_{k-1}}$ is a failure state. Note that if
$M_0=\emptyset$, $\hat{s_0}$ is a failure state. To check whether or
not a finite counterexample is spurious, $M_0$, $M_1$, $M_2$, ...
are computed in turn until the first state $\hat{s_k}$ where
$M_k=\emptyset$ is found, or the last state in the counterexample is
reached.

For  infinite counterexamples, it is more complicated to be dealt
with since the last state in the counterexample can never be
reached. Thus, a polynomial number of unwinding of the loop in the
counterexample is needed \cite{CGJLV00}. That is an infinite
counterexample can be reduced to a finite counterexample by
unwinding the loop for a polynomial number of times. Accordingly,
{\scshape SplitPath} can be used again to check whether or not this
infinite counterexample is spurious.

\section{Failure States and Spurious Counterexamples} In
\cite{Clarke04,HSHGS10}, a spurious counterexample is informally
defined by: a counterexample in the abstract model which does not
exist in the concrete model. In this section, we give a formal
definition for spurious counterexamples based on the the formal
definition of failure states.

\begin{figure}[htp]
\centerline{\includegraphics[height=4.0cm]{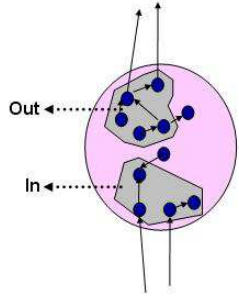}}\caption{$In_{\hat{s_i}}$
and $Out_{\hat{s_i}}$}\label{fig:inout}
\end{figure}

To this end, $In_{\hat{s_i}}^0$, $In_{\hat{s_i}}^1$, ...,
$In_{\hat{s_i}}^n$ and $In_{\hat{s_i}}$ are defined first:
$$
\begin{array}{llll}
In_{\hat{s_i}}^0&=&\{s\mid s\in h^-(\hat{s_i}), s'\in h^-(\hat{s_{i-1}}) \mbox{ and }(s',s)\in R \}\\

In_{\hat{s_i}}^1&=&\{s\mid s\in h^-(\hat{s_i}), s'\in In_{\hat{s_i}}^0 \mbox{ and }(s',s)\in R \}\\
&...&\\

In_{\hat{s_i}}^n&=&\{s\mid s\in h^-(\hat{s_i}), s'\in In_{\hat{s_i}}^{n-1} \mbox{ and }(s',s)\in R \}\\
&...&\\

In_{\hat{s_i}}&=&\bigcup\limits_{i=0}^\infty In_{\hat{s_i}}^i
\end{array}
$$
Clearly, $In_{\hat{s_i}}^0$ denotes the set of states in
$h^-(\hat{s_i})$ with inputting edges from the states in
$h^-(\hat{s_{i-1}})$, and $In_{\hat{s_i}}^1$ stands for the set of
states in $h^-(\hat{s_i})$ with inputting edges from the states in
$In_{\hat{s_i}}^0$, and $In_{\hat{s_i}}^2$ means the set of states
in $h^-(\hat{s_i})$ with inputting edges from the states in
$In_{\hat{s_i}}^1$, and so on. Thus, $In_{\hat{s_i}}$ denotes the
set of states in $h^-(\hat{s_i})$  that are reachable from some
state in $h^-(\hat{s_{i-1}})$ as illustrated in the lower gray part
in Fig. \ref{fig:inout}. Note that there must exist a natural number
$n$, such that $\bigcup\limits_{i=0}^{n+1}In_{\hat{s_i}}^{i}
=\bigcup\limits_{i=0}^{n}In_{\hat{s_i}}^{i}$ since $h^-(\hat{s_i})$
is finite. Note that for state $\hat{s_0}$,
$$
\begin{array}{llll}
In_{\hat{s_0}}^0&=&\{s\mid s\in (h^-(\hat{s_0})\cap I) \}\\

In_{\hat{s_0}}^1&=&\{s\mid s\in h^-(\hat{s_0}), s'\in In_{\hat{s_0}}^0 \mbox{ and }(s',s)\in R \}\\
&...&\\
\end{array}
$$
That is only $In_{\hat{s_0}}^0$ is defined differently since there
are no pre states.

Similarly,  $Out_{\hat{s_i}}^0$, $Out_{\hat{s_i}}^1$, ...,
$Out_{\hat{s_i}}^n$  and $Out_{\hat{s_i}}$ can also be defined.

$$
\begin{array}{llll}
Out_{\hat{s_i}}^0&=&\{s\mid s\in h^-(\hat{s_i}), s'\in h^-(\hat{s_{i+1}}) \mbox{ and }(s,s')\in R \}\\

Out_{\hat{s_i}}^1&=&\{s\mid s\in h^-(\hat{s_i}), s'\in Out_{\hat{s_i}}^0 \mbox{ and }(s,s')\in R \}\\
&...&\\
\end{array}
$$

$$
\begin{array}{llll}
Out_{\hat{s_i}}^n&=&\{s\mid s\in h^-(\hat{s_i}), s'\in Out_{\hat{s_i}}^{n-1} \mbox{ and }(s,s')\in R \}\\
&...&\\

Out_{\hat{s_i}}&=&\bigcup\limits_{i=0}^\infty Out_{\hat{s_i}}^i\\
\end{array}
$$
Where $Out_{\hat{s_i}}^0$ denotes the set of states in
$h^-(\hat{s_i})$ with outputting edges to the states in
$h^-(\hat{s_{i+1}})$, and $Out_{\hat{s_i}}^1$ stands for the set of
states in $h^-(\hat{s_i})$ with outputting edges to the states in
$Out_{\hat{s_i}}^0$, and $Out_{\hat{s_i}}^2$ means the set of states
in $h^-(\hat{s_i})$ with outputting edges to the states in
$Out_{\hat{s_i}}^1$, and so on. Thus, $Out_{\hat{s_i}}$ denotes the
set of states in $h^-(\hat{s_i})$ from which some state in
$h^-(\hat{s_{i+1}})$ are reachable as depicted in the higher gray
part in Fig. \ref{fig:inout}. Similar to $In_{\hat{s_i}}$, there
must exist a natural number $n$, such that
$\bigcup\limits_{i=0}^{n+1}Out_{\hat{s_i}}^{i}
=\bigcup\limits_{i=0}^{n}Out_{\hat{s_i}}^{i}$. Note that for the
last state $\hat{s_n}$ in a finite counterexample,
$$
\begin{array}{llll}
Out_{\hat{s_n}}^0&=&\{s\mid s\in h^-(\hat{s_n})\cap F \}\\

Out_{\hat{s_n}}^1&=&\{s\mid s\in h^-(\hat{s_n}), s'\in Out_{\hat{s_n}}^0, \mbox{ and }(s,s')\in R \}\\
&...&
\end{array}
$$
where $F$ is the set of states without any successors in the
original model.

Accordingly, a failure state can be defined as follows.
\begin{Def}{\bf (Failure States) }\rm A state $\hat{s_i}$ in a counterexample $\hat{\Pi}$ is a failure state if, and only if $In_{\hat{s_i}}\cap Out_{\hat{s_i}}=\emptyset$. \hfill{$\Box$}
\end{Def}

Further, given a failure state $\hat{s_i}$ in a counterexample
$\hat\Pi$, the set of the origins of $\hat{s_i}$, $h^-(\hat{s_i})$,
is separated into three sets, $\mathcal{D}=In_{\hat{s_i}}$ (the set
of dead states), $\mathcal{B}=Out_{\hat{s_i}}$ (the set of bad
states) and $\mathcal{I}=h^-(\hat{s_i})\setminus(\mathcal{D}\cup
\mathcal{B})$ (the set of the isolated states).

\begin{Def}{\bf (Spurious Counterexamples) }\rm A counterexample $\hat{\Pi}$ in an abstract model $\hat{K}$ is spurious
if there exists at least one failure state $\hat{s_i}$ in
$\hat{\Pi}$ \hfill{$\Box$}
\end{Def}

\begin{Exm}\rm
Fig. \ref{fig:spurious} shows a spurious counterexample where state
$\hat{2}$ is a failure  state.

In the set, $h^-(\hat{2})=\{7,8,9\}$, of the origins of state
$\hat{2}$, $9$ is  a dead state, $7$ is a bad state, and $8$ is an
isolated state. \begin{figure}[htp]
\centerline{\includegraphics[height=2.8cm]{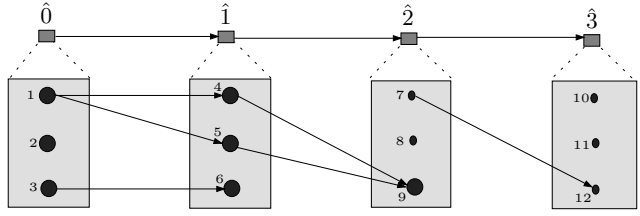}}\caption{A
Spurious Path}\label{fig:spurious}
\end{figure}\hfill{$\Box$}
\end{Exm}

\section{Algorithms for Detecting Spurious Counterexamples}

Based on the formal definition of spurious counterexample, new
algorithms for checking whether or not a counterexample is spurious
are presented in this section.

\subsection{Algorithm by Detecting the First Failure State}

Algorithm {\scshape CheckSpurious-I} takes a counterexample as input
and outputs the first failure state in the counterexample. Note that
a counterexample may be a finite path $<s_0,s_1,...,s_n>$, $n\geq
0$, or an infinite path $<s_0,s_1,...,(s_i,...,s_j)^\omega>$, $0\leq
i\leq j$, with a loop suffix (a suffix produced by a loop). For the
finite one, it can be checked directly; while for an infinite one,
we need only to check its Complete Finite Prefix (CFP)
$<s_0,s_1,...,s_i,...,s_j>$ since whether or not a state $s_i$ is a
failure state only relies on its pre and post states. It is pointed
out that in the CFP $<s_0,s_1,...,s_i,...,s_j>$ of an infinite
counterexample,
$$
\begin{array}{llll}
Out_{\hat{s_j}}^0&=&\{s\mid s\in h^-(\hat{s_j}), s'\in h^-(\hat{s_{i}}) \mbox{ and }(s,s')\in R \}\\

Out_{\hat{s_j}}^1&=&\{s\mid s\in h^-(\hat{s_j}), s'\in Out_{\hat{s_j}}^0 \mbox{ and }(s,s')\in R \}\\
&...&\\
\end{array}
$$
since the post state of $\hat{s_j}$ is $\hat{s_i}$.

{\small\begin{algorithm}[h] \caption{: {\scshape
CheckSpurious-I}($\hat{\Pi}$)}
  {\bf Input}: a counterexample $\hat{\Pi}=<\hat{s_0},\hat{s_1},...,\hat{s_n}>$ in the abstract model $\hat{K}=(\hat{S},\hat{S_0},\hat{R},\hat{L})$, and the original model $K=(S,S_0,R,L)$\\
  {\bf Output}: a failure state $s_f$
    \begin{algorithmic}[1]
\STATE {\bf Initialization}: $int$ $i=0$;
        \WHILE {$i\leq n$}
            \STATE  {\bf if} $In_{\hat{s_i}}\cap Out_{\hat{s_i}}\not=\emptyset$, $i=i+1$;
            \STATE  {\bf else} return $s_f=\hat{s_{i}}$; break;
        \ENDWHILE \label{code:recentEnd}
\STATE {\bf if} $i==n+1$, return $\hat{\Pi}$ is a real
counterexample;
    \end{algorithmic}
\end{algorithm}}

\paragraph{\bf Algorithm Analyzing} In algorithm {\scshape CheckSpurious-I}, to check whether or not a
state $\hat{s_i}$ is a failure state only relies on $\hat{s_i}$'s
pre and post states, $\hat{s_{i-1}}$ and $\hat{s_{i+1}}$; while in
algorithm {\scshape SplitPath}, to check state $\hat{s_i}$ is up to
the prefix, $\hat{s_0},...,\hat{s_{i-1}}$, of $\hat{s_i}$. Based on
this, to check a periodic infinite counterexample, several
repetitions of the periodic parts are needed in {\scshape
SplitPath}. In contrast, this can be easily done by checking the
complete finite prefix  $<s_1,s_2,...,s_i,...,s_j>$ in algorithm
{\scshape CheckSpurious-I}. Thus, the polynomial number of unwinding
of the loop can be avoided. That is for infinite counterexamples,
the finite prefix to be checked will be polynomial shorter than the
one in algorithm {\scshape SplitPath}.

\subsection{Algorithm by Detecting the Heaviest Failure State}
\begin{figure}[htp]
\centerline{\includegraphics[height=2.0cm]{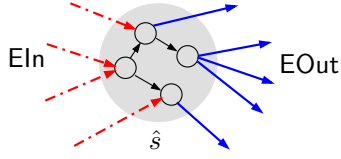}}\caption{In and
out edges}\label{fig:ioe}
\end{figure}

In algorithm {\scshape SplitPath} and {\scshape CheckSpurious-I},
always, the first failure state is detected. Then further refinement
will be done based on the analysis of this failure state.
{\small\begin{algorithm}[h] \caption{: {\scshape
CheckSpurious-II}($\hat{\Pi}$)}
  {\bf Input}: a counterexample $\hat{\Pi}=<\hat{s_0},\hat{s_1},...,\hat{s_n}>$ in the abstract model $\hat{K}=(\hat{S},\hat{S_0},\hat{R},\hat{L})$, and the original model $K=(S,S_0,R,L)$\\
  {\bf Output}: the heaviest failure state $s_f$
    \begin{algorithmic}[1]
        \STATE{\bf Sorting:} the heavier the earlier (stored in array \emph{int} $w[n+1]$);
        \STATE {\bf Initialization}: $int$ $i=0$;
        \WHILE {$i\leq n$}
            \STATE  {\bf if} $In_{\hat{s_{w[i]}}}\cap Out_{\hat{s_{w[i]}}}\not=\emptyset$, $i=i+1$;
            \STATE  {\bf else} return $s_f=\hat{s_{{w[i]}}}$; break;
        \ENDWHILE \label{code:recentEnd}
\STATE {\bf if} $i==n+1$, return $\hat{\Pi}$ is a real
counterexample;
    \end{algorithmic}
\end{algorithm}}
Possibly,
several failure states may occur in one counterexample, so which one
is chosen to be refined is not considered in {\scshape SplitPath}.
Obviously, if a failure state shared by more paths is refined, a
number of model checking iterations will be saved in the whole
abstract-refinement loop. With this consideration, we will check the
states which is shared by more paths first. To do so, for an
abstract state $\hat{s}$ as illustrated in Fig.\ref{fig:ioe},
$EIn(\hat{s})$ and $EOut(\hat{s})$ are defined. $EIn(\hat{s})$
equals to the number of edges connecting to the states in
$h^-(\hat{s})$ from the states outside of $h^-(\hat{s})$; and
$EOut(\hat{s})$ is the number of edges connecting to the states out
of $h^-(\hat{s})$ from the states in $h^-(\hat{s})$. Accordingly,
$EIn(\hat{s})\times EOut(\hat{s})$ is the number of the paths where
$\hat{s}$ occurs. For convenience, we call $EIn(\hat{s})\times
EOut(\hat{s})$ the weight of the abstract state $\hat{s}$. Based on
this, algorithm {\scshape CheckSpurious-II} is given for detecting
the heaviest failure state in a counterexample. In {\scshape
CheckSpurious-II}, an array \emph{w[i]} is used to store the indexes
of the states in the counterexample by the heavier the earlier.

\subsection{Parallel Algorithms}
Considering whether or not a state $\hat{s_i}$ is a failure state
only relies on the pre- and post- states, $\hat{s_{i-1}}$ and
$\hat{s_{i+1}}$, of $\hat{s_i}$, the algorithm can be naturally
paralleled as presented in algorithm {\scshape CheckSpurious-III}
and {\scshape CheckSpurious-IV}.

In {\scshape CheckSpurious-III}, anytime, if a failure state is
detected by a processor, all the processors will be stop and the
failure state is returned. Otherwise, if no failure states are
reported, the counterexample is a real one. That is the algorithm
always reports the first detected failure state obtained by the
processors. Note that a boolean array $c[n]$ is used to indicate
whether or not a state in the counterexample is a failure one.
Initially, for all $0\leq i\leq n$, $c[i]$ is undefined
($c[i]=\bot$). $c[i]==true$ means state $\hat{s_i}$ is not a failure
state.

{\small\begin{algorithm}[h] \caption{: {\scshape
CheckSpurious-III}($\hat{\Pi}$)}
  {\bf Input}: a counterexample $\hat{\Pi}=<\hat{s_0},\hat{s_1},...,\hat{s_n}>$ in the abstract model $\hat{K}=(\hat{S},\hat{S_0},\hat{R},\hat{L})$, and the original model $K=(S,S_0,R,L)$ in shared memory\\
$n$: the number of processors\\
$k$: processor id\\
  {\bf Output}: a failure state $s_f$
    \begin{algorithmic}[1]
    \STATE {\bf Initialization}: $bool$ $c[n+1]=\{\bot,...,\bot\}$;
        \FOR {$k=0$ to $n$ do in parallel}
            \STATE  {\bf if} $In_{\hat{s_k}}\cap Out_{\hat{s_k}}\not=\emptyset$, $c[k]=ture$;
            \STATE  {\bf else} return $s_f=\hat{s_{k}}$; stop all processors;
        \ENDFOR \label{code:recentEnd}
\STATE {\bf if} for all $0\leq i\leq n$, $c[i]==ture$, return
$\hat{\Pi}$ is a real counterexample;
    \end{algorithmic}
\end{algorithm}}

In {\scshape CheckSpurious-IV}, the weight of the states are
considered, and always the heaviest failure state is found.

 {\small\begin{algorithm}[h]
\caption{: {\scshape CheckSpurious-IV}($\hat{\Pi}$)}
  {\bf Input}: a counterexample $\hat{\Pi}=<\hat{s_0},\hat{s_1},...,\hat{s_n}>$ in the abstract model $\hat{K}=(\hat{S},\hat{S_0},\hat{R},\hat{L})$, and the original model $K=(S,S_0,R,L)$ in shared memory\\
$n$: the number of processors\\
$k$: processor id\\
  {\bf Output}: a failure state $s_f$
    \begin{algorithmic}[1]
      \STATE{\bf Sorting:} the heavier state first (stored in array \emph{int} $w[n+1]$);
        \FOR {$k=0$ to $n$ do in parallel}
            \STATE  {\bf if} $In_{\hat{s_k}}\cap Out_{\hat{s_k}}\not=\emptyset$, $c[k]=ture$;
            \STATE  {\bf else} return $c[k]=false$;
        \ENDFOR \label{code:recentEnd}
\STATE {\bf if} for all $1\leq i\leq n$, $c[i]=ture$, return
$\hat{\Pi}$ is a real counterexample;

 \STATE {\bf else} return $s_f=s_i$ such that $c[i]==false$, and for
 any state $\hat{s_j}$, if the weight of $\hat{s_j}$ is heavier than $\hat{s_i}$, $c[j]==true$;
    \end{algorithmic}
\end{algorithm}}

\section{Conclusion}
Based on a formal definition of spurious paths, a novel approach for
detecting spurious counterexamples are presented in this paper. In
the new approach, whether or not a state $\hat{s_i}$ is a failure
state only relies on $\hat{s_i}$'s pre- and post- states in the
counterexample. So, for infinite counterexample, the polynomial
number of unwinding of the loop can be avoided. Further, the
algorithm can be easily improved by detecting the heaviest failure
state such that a number of model checking iterations can be saved
in the whole abstract-refinement loop. Also, the algorithm can be
naturally parallelled.

The presented algorithms are useful in improving the abstract based
model checking, especially the counterexample guided abstraction
refinement model checking. In the near future, the proposed
algorithm will be implemented and integrated into the tool CEGAR.
Further, some case studies will be conducted to evaluate
  the algorithms.

\end{document}